\documentclass[preprint]{revtex4}

\usepackage{graphicx}
\usepackage{color}
\usepackage{dcolumn}
\usepackage{bm}
\usepackage{subfig}

\def\be{\begin{equation}}
\def\ee{\end{equation}}
\def\bea{\begin{eqnarray}}
\def\eea{\end{eqnarray}}

\def\calh{\mathcal{H}}
\def\calp{\mathcal{P}}

\begin{document}

\title{Timelike Vector Field Dynamics in the Early Universe}
\author{Seoktae Koh} \email[email: ]{skoh@itp.ac.cn}
\author{Bin Hu} \email[email: ]{hubin@itp.ac.cn}

\address{Institute of Theoretical Physics, Chinese Academy of Sciences,
P.O. Box 2735, Beijing, 100080, China}


\begin{abstract}
We study the dynamics of a timelike vector field 
when the background spacetime is in an accelerating phase in the
early universe.
It is shown that a timelike vector field is difficult to
realize an inflationary phase, so we investigate the
evolution of a vector field within a scalar field driven inflation model.  
And we calculate the power spectrum of the vector field without considering
the metric perturbations. While the time component of 
the vector field perturbations provides 
a scale invariant spectrum when $\xi = 0$
, where $\xi$ is a nonminimal coupling parameter, both the longitudinal and
transverse perturbations 
give a scale invariant spectrum when $\xi = 1/6$
in the absence of the coupling terms. The deviation
of the power spectrum due to the coupling terms are calculated
 by use of the Greens' function.

\end{abstract}

\maketitle

\section{Introduction} \label{intro}

A recent 
remarkable development in observational data reveals
many interesting, unusual features 
which could not be predicted theoretically before
such as the present accelerating phase  by the 
supernova data \cite{Perlmutter:1998np},
the suppression of the cosmic microwave background (CMB) angular
 power in the low multipole moments \cite{Bennett:1996ce} and
the large scale anomaly in CMB \cite{Eriksen:2003db}.
These progresses enable us to enter into
the precision cosmology theoretically and to seek the new physics.
Recently, vector fields are widely investigated to explore 
inflation \cite{Ford:1989me, Golovnev:2008cf, Kanno:2008gn, Kanno:2006ty}
and also to explain the dark energy problem 
\cite{ArmendarizPicon:2004pm, Jimenez:2008au, Koivisto:2008xf,
 Boehmer:2007qa}.

It was known to be difficult to realize a vector field inflation model
because the effective mass of a vector field must be order of the Hubble scale
\cite{Ford:1989me}. But recently a successful inflation model is 
achieved in \cite{Golovnev:2008cf}.
They have used either a triad (a triplet of orthogonal vector fields) 
\cite{ArmendarizPicon:2004pm} or large $N$ vector fields 
for an isotropic spacetime and taken into account 
a nonminimal coupling for a slow-roll phase.
In that case, vector field inflation looks like
a scalar field chaotic inflation model.
A vector field can also play against 
the cosmic no hair theorem in anisotropic inflation \cite{Kanno:2008gn}. 
In a usual scalar field driven 
inflation model, even if the initial stage starts off from an anisotropic 
background spacetime ({\it e.g.} Bianchi type models), 
the anisotropy would disappear very soon because of 
an accelerating expansion. But if there exists a vector field,
the anisotropy will remain even if the universe undergoes
a period of inflation. But the vector field models 
\cite{Ford:1989me, Golovnev:2008cf, Kanno:2008gn}
which have the standard Maxwell kinetic term are known to be unstable 
\cite{Himmetoglu:2008zp} because they contain a ghost.
The presence of the ghost is due to the sign that one
needs to choose for the nonminimal coupling of the vector field to the
curvature.

Although a vector field can  provide some interesting properties
in inflation or
in the dark energy problem,
it is difficult to handle, especially with the linear perturbations
\cite{ArmendarizPicon:2004pm}.
Because a non-vanishing vector field
breaks spatial isotropy in a background spacetime and further
makes it impossible to decompose the perturbation modes \cite{Bardeen:1980kt}
- scalar, vector and tensor perturbation - in the linear perturbation
theory.
 One way to resolve spatial anisotropy
is to use a triad \cite{ArmendarizPicon:2004pm} or 
large $N$ random
vector fields \cite{Golovnev:2008cf}. The other way is to 
use an anisotropic background spacetime 
\cite{Ford:1989me, Kanno:2008gn,Dimopoulos:2006ms}.
The difficulty, however, in using decomposition theorem in the linear
perturbations causes new obstacles to the calculation of the power spectrum
and to the fitting with the observational data.
In spite of this difficulty in the linear perturbations,
the gravitational wave spectrum is calculated in \cite{Golovnev:2008hv},
in which they used the underlying symmetry 
in order to eliminate the mode coupling terms.
The power spectra of scalar and tensor perturbations of a
timelike vector fields are calculated in \cite{Lim:2004js} with a
fixed-norm condition. 
And in \cite{Dimopoulos:2006ms}, the power spectrum of
the longitudinal and transverse component of a spacelike vector field
are calculated,
but the gravitational metric perturbations were not considered.
While the longitudinal component is scale invariant
for $m^2 \ll H^2$, the transverse component is scale invariant
when a vector field is coupled to the gravity nonminimally ($\xi =1/6$)
for $m^2 \ll H^2$.

In order to avoid the breaking of a spatial isotropic background and
the existence of mode coupling between different perturbation modes,
we  will take into account  a timelike vector field instead of
a spacelike vector field \cite{Ford:1989me, Kanno:2008gn, Golovnev:2008cf}.
Since the Maxwell kinetic energy, $-F_{\mu\nu}F^{\mu\nu}/4$,
could not present any dynamics 
for a timelike vector field, we will consider a vector field Lagrangian
with the general kinetic energy terms
\cite{Carroll:2004ai, Lim:2004js} 
for the nontrivial dynamics of a timelike vector field
\bea
\mathcal{L}_A = -\frac{1}{2}\beta_1 \nabla_{\mu} A_{\nu}
\nabla^{\mu}A^{\nu} -\frac{1}{2}\beta_2 (\nabla_{\rho}A^{\rho})^2
-\frac{1}{2}\beta_3 \nabla_{\mu}A^{\nu}\nabla_{\nu} A^{\mu}
-\frac{1}{2}(m^2 - \xi R)A_{\mu}A^{\mu}.
\eea
In this paper, especially 
we will only focus on  $\beta_1 = \beta_2 = -\beta_3 = 1$ case,
then the Lagrangian for a vector field can be written as
\bea
\mathcal{L}_A = -\frac{1}{4} F_{\mu\nu}F^{\mu\nu} 
- \frac{1}{2}(\nabla_{\mu}A^{\mu})^2
- \frac{1}{2}(m^2 -\xi R)A_{\mu}A^{\mu},
\eea
where $F_{\mu\nu} = \nabla_{\mu}A_{\nu} - \nabla_{\nu}A_{\mu}$.
Unlike in \cite{Carroll:2004ai, Lim:2004js}, 
we do not require a fixed-norm condition, $A_{\mu}A^{\mu}  
=-m^2$.

This paper is organized as follows: in Section \ref{background}
we describe our model and discuss about the difficulty in realization 
of successful inflation with a timelike vector field. And 
we calculate 
the evolution of a vector field in a scalar field driven
inflation model. In Section \ref{pert} we calculate
the linear perturbation of a vector field without considering 
metric perturbations on a scalar field driven inflationary background.
And the power spectrum of a vector field is calculated.
We discuss about the spectral index of the vector field perturbations
 and briefly comment about
the linear perturbations including gravitational metric perturbations.
We conclude in Section \ref{conclusion}.

\section{Background dynamics with timelike vector fields}
\label{background}

We start  with an action of a massive vector field which is coupled 
nonminimally to gravity
\bea
S = \int d^4 x\sqrt{-g} \left[\frac{1}{16\pi G} R
-\frac{1}{4} F_{\mu\nu}F^{\mu\nu} - \frac{1}{2}(\nabla_{\mu}A^{\mu})^2
-\frac{1}{2}(m^2 - \xi R)A_{\mu}A^{\mu}
\right] 
\label{actionv}
\eea
where $\xi$ is a nonminimal coupling parameter.
This nonminimal coupling
term can make it possible to occur a successful inflationary period with
a spacelike vector field \cite{Golovnev:2008cf}.

By varying the action (\ref{actionv}) with respect to $A_{\mu}$,
one obtains the equations of motion,
\bea
\nabla_{\mu}F^{\mu\nu} +\nabla^{\nu}\nabla_{\rho}A^{\rho} 
-(m^2-\xi R)A^{\nu} = 0.
\label{eom}
\eea
And  Einstein equations can be obtained by varying the action with
respect to $g_{\mu\nu}$
\bea
R_{\mu\nu} &-& \frac{1}{2}g_{\mu\nu} R = 8\pi G T_{\mu\nu}^{(A)}, \\
T_{\mu\nu}^{(A)} &=& {F_{\mu}}^{\rho}F_{\nu\rho} -\frac{1}{4}g_{\mu\nu}
F_{\rho\sigma}F^{\rho\sigma} 
-A_{\mu}\nabla_{\nu}\nabla_{\rho}A^{\rho} -A_{\nu}\nabla_{\mu}
\nabla_{\rho}A^{\rho} \nonumber \\
& &+ \frac{1}{2}g_{\mu\nu}(\nabla_{\rho}A^{\rho})^2
+g_{\mu\nu}A^{\rho}\nabla_{\rho}\nabla_{\sigma}A^{\sigma}
+(m^2-\xi R)A_{\mu}A_{\nu}  \nonumber \\
& &-\xi R_{\mu\nu}A_{\rho}A^{\rho} 
+\xi(\nabla_{\mu}\nabla_{\nu} - g_{\mu\nu}\nabla^2)A_{\rho}A^{\rho}
-\frac{1}{2}g_{\mu\nu}(m^2 - \xi R)A_{\rho}A^{\rho}.
\eea

Since we want a homogeneous and
isotropic background spacetime, we 
consider a timelike vector field, $A_{\mu}A^{\mu} < 0$, and choose
$A_{\mu} = (\chi, \vec{0})$. 
In the spatially flat  FRW metric
\bea
ds^2 = -dt^2 + a^2(t) \delta_{ij} dx^idx^j,
\label{metric}
\eea
the Einstein equations and  equation of motion for $\chi$
can be expressed as
\bea
H^2 &=& \frac{8\pi G}{3}\rho_{\chi}  \nonumber \\
&=& \frac{8\pi G}{3} \Biggr[
-\frac{1}{2}\dot{\chi}^2 -3(1-2\xi)H\chi \dot{\chi}
-12\xi \dot{H}\chi^2
-\frac{3}{2}(3+14\xi)H^2 \chi^2 
+\frac{3}{2}m_{\chi}^2 \chi^2  \Biggl],  \label{vector_ein1}\\
\dot{H} &=& -4\pi G(\rho_{\chi} + p_{\chi})  \nonumber \\
&=& -4\pi G\Biggr[ (1+2\xi)m_{\chi}^2 \chi^2
-2\xi (\dot{\chi}^2 -5H\chi \dot{\chi}  
+(1+6\xi)\dot{H}\chi^2
+6(1+2\xi)H^2 \chi^2) \Biggl], \label{vector_ein2}\\
\ddot{\chi} &+& 3H\dot{\chi} + \left[
3(1 - 2\xi) \dot{H} - 12\xi
H^2 + m_{\chi}^2\right]\chi = 0,
\label{vectorf_eom}
\eea
where we have used $R = 6(\dot{H}+2H^2)$. 

We need to check if the timelike vector field could
generate an accelerating phase. If we define $m_{eff}^2
= 3(1-2\xi)\dot{H}-12\xi H^2 +m_{\chi}^2$ in (\ref{vectorf_eom}),
it is required $ m_{eff}^2 \ll H^2$  for a slow-rolling field
and $ \dot{H}\ll H^2$ for sufficient inflation,
which means $N_e =\int Hdt \geq 50$ to fit to the observational data.
From the expression in (\ref{vector_ein1}), the energy density of
the vector field is not positive definite. 
The vector field may have a negative energy density
in some range. So we need to constrain on $\chi$ to avoid
a negative energy density in our discussion.

If we assume that the vector field can generate inflation,
one obtains from (\ref{vector_ein1}) for $\xi=0$
\bea
H^2 \simeq \frac{8\pi}{3m_{pl}^2}\left(1
+12\pi \left(\frac{\chi}{m_{pl}}\right)^2\right)^{-1}
\left[-4\pi m_{\chi}^2\left(\frac{\chi}{m_{pl}}\right)^2
+\frac{5}{2}m_{\chi}^2 \right] \chi^2,
\label{pos_energy}
\eea 
where we have neglected the kinetic energy term and 
 used (\ref{vectorf_eom}) in which we neglect $\ddot{\chi}$.
But we keep $\dot{H}$ term  when we obtain (\ref{pos_energy})
because it has the same order of magnitude as the potential term
from (\ref{vector_ein2}) for $\xi=0$.
Here $m_{pl}$ is Planck mass.
In order to guarantee the positive energy density,
$\chi$ should be  constrained by $\chi \ll m_{pl}$.
This is different from a scalar field chaotic inflation model in 
which the initial amplitude of an inflaton 
should be larger than the Planck mass
to have a sufficient inflationary period. 
 
As long as $\chi \ll m_{pl}$, $H^2 \simeq \dot{H}$ and $H^2 < m_{eff}^2$
where we have used $\dot{H} \sim -4\pi m_{\chi}^2 
\left(\frac{\chi}{m_{pl}}\right)^2$ when $\xi = 0$.
These conditions contradict with those for sufficient inflation.
So it is hard to realize inflation by the
timelike vector field.
 Even if the nonminimal coupling is taken into account,
$\dot{H}$ is shown
to be of the same order of $H^2$  and the potential of
the vector field in (\ref{vector_ein1}) and (\ref{vector_ein2})
\bea
H^2 &\sim& \frac{8\pi}{3m_{pl}^2} \left[3(1-8\xi+4\xi^2)\dot{H}\chi^2
+\frac{1}{2}(5-4\xi)m^2 \chi^2\right], \\
\dot{H} &\sim& -\frac{4\pi}{m_{pl}^2}\left[
4\xi(4\xi -3) H^2 \chi^2 +\frac{1}{3}(3-4\xi)m^2 \chi^2\right],
\eea
where we have used $\chi \ll m_{pl}^2$.
As a result,
the slow-roll conditions could not be fulfilled.

\begin{figure}
\includegraphics[width=0.6\textwidth]{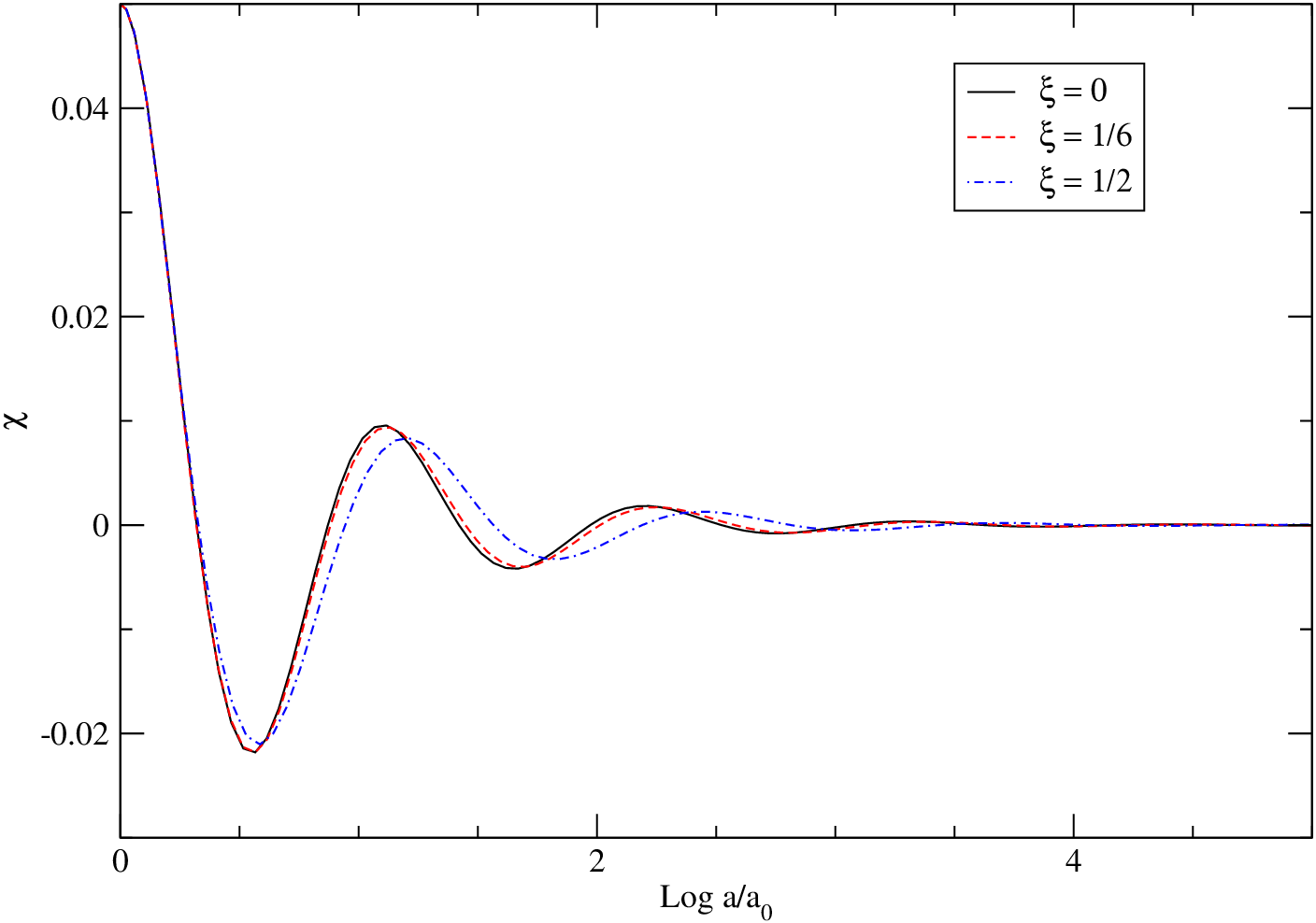}
\caption{The evolution of $\chi$ is plotted 
with a different nonminimal coupling parameter $\xi$.}
\label{fig1a}
\end{figure}

We calculate (\ref{vector_ein1}), (\ref{vector_ein2}) and 
(\ref{vectorf_eom}) numerically
to confirm these analytical arguments. The results are shown in 
Fig. \ref{fig1a}. We set to $\chi_i = 0.05 m_{pl}$ and $\dot{\chi} = 0$
as an initial
condition. As expected,
we could not obtain the inflationary solutions. Even if we consider the
nonminimal coupling $\xi$, it could not help to get an accelerating
phase. 
We consider $\xi = 1/6$ and $1/2$ in Fig. \ref{fig1a} but 
it does not improve the results.

\begin{figure}
\includegraphics[width=0.6\textwidth]{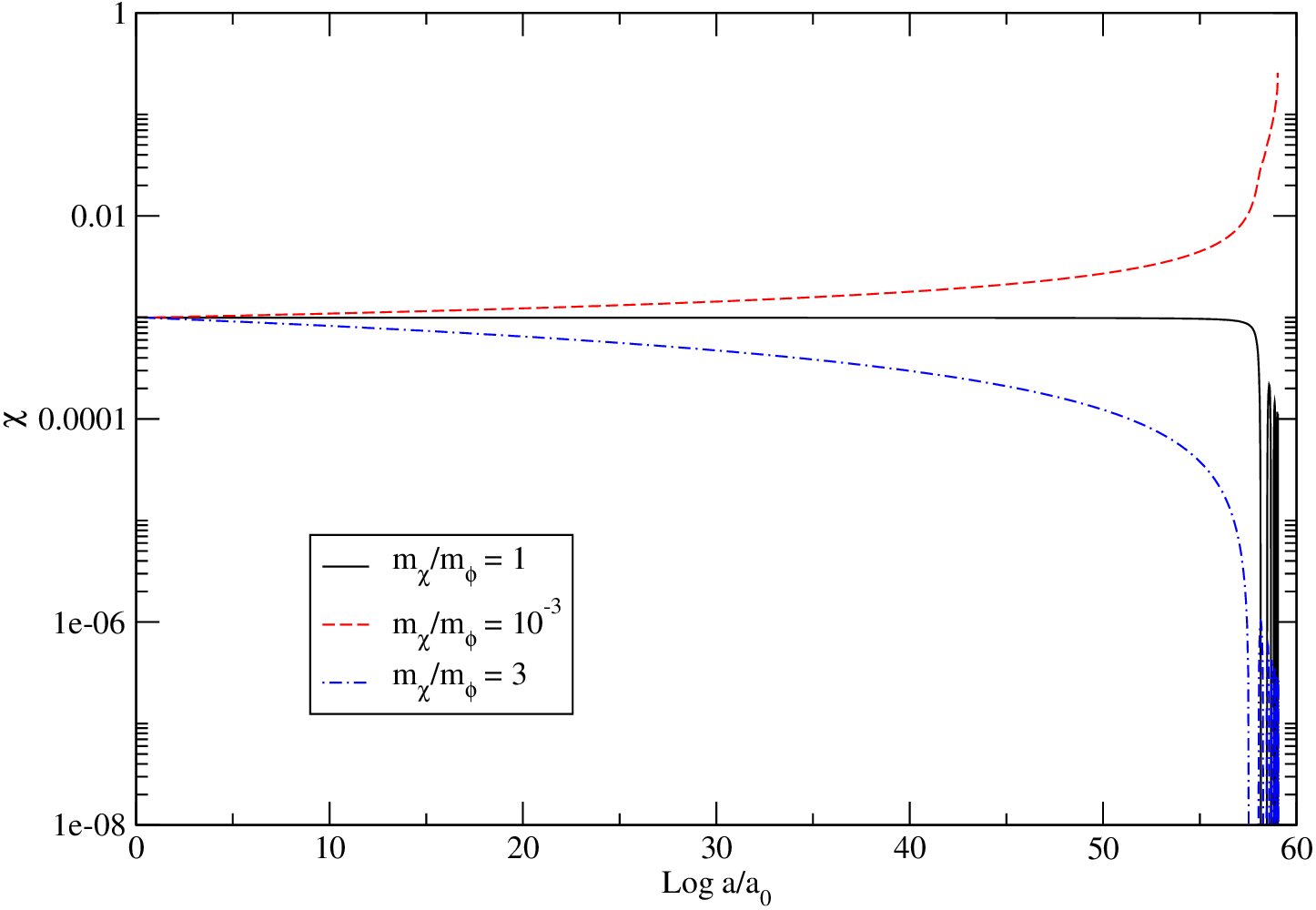}
\caption{The evolution of $\chi$ is plotted depending on $m_{\chi}/m_{\phi}$.
Here a scalar field $\phi$ is responsible for an accelerating expansion.
We have used $m_{\phi}/m_{pl} = 10^{-5},~ \phi_i = 
3m_{pl}$ and $\chi_i = 10^{-3}m_{pl}$.}
\label{fig1b}
\end{figure}

Even if the vector field have failed to generate inflation,
it may play a role as a curvaton \cite{Dimopoulos:2006ms, Lyth:2001nq}
which can generate the curvature 
perturbation after the end of inflation. So we need to investigate 
the dynamics of the vector field during an inflationary period which 
occurs due to an additional matter such as a scalar field.
Then we add a scalar field $\phi$, which drives an accelerating phase,
  to the vector field action:
\bea
S &=& \int d^4 x \sqrt{-g} \Biggl[\frac{1}{16\pi G} R
-\frac{1}{2}\partial_{\mu}\phi \partial^{\mu}\phi
-V(\phi) 
-\frac{1}{4} F_{\mu\nu}F^{\mu\nu}  \nonumber \\
& &~~~~~~~~~~~~
- \frac{1}{2}(\nabla_{\mu}A^{\mu})^2 
-\frac{1}{2}(m_{\chi}^2 - \xi R)A_{\mu}A^{\mu}
\Biggr],
\label{spec_action}
\eea
where we will consider a massive scalar field potential $V(\phi) = \frac{1}{2}
m_{\phi}^2 \phi^2$ in the present paper.

One obtains the equations of motion for $\phi$ and $\chi$
\bea
& &\ddot{\phi} + 3H \dot{\phi} + V_{,\phi} = 0, \label{scalar_eom}\\
& &\ddot{\chi} +3H\dot{\chi} + \left[
3(1 - 2\xi) \dot{H} - 12\xi
H^2 + m_{\chi}^2\right]\chi = 0 \label{vector_eom2},
\eea
where the Einstein equations become
\bea
H^2 &=&  \frac{8\pi G}{3} \left(\frac{1}{2}\dot{\phi}^2 + V(\phi)
+\rho_{\chi}\right) \simeq \frac{8\pi G}{3} V(\phi), \\
\dot{H} &=& -4\pi G(\dot{\phi}^2 +\rho_{\chi}+p_{\chi}),
\eea
and $\rho_{\chi}$ and $p_{\chi}$ are given in (\ref{vector_ein1})
and (\ref{vector_ein2}).

Since the scalar field is responsible for an inflationary phase, we
can assume $H \simeq {\rm const.}$ and use the slow-roll conditions,
 $|\dot{H}|\ll H^2,~~ \frac{1}{2}\dot{\phi}^2 \ll V(\phi)$ and
$\ddot{\phi} \ll 3H\dot{\phi}$ . In addition, in order 
for the scalar field $\phi$ to be a dominant component, we assume
$\rho_{\chi} \ll \rho_{\phi}$. 

Then one can easily get the solution of 
(\ref{scalar_eom}) for $V = \frac{1}{2}m^2\phi^2$
\bea
\phi(t) = \phi_i - \frac{m_{\phi}}{\sqrt{12\pi G}}(t-t_i),
\eea
where $\phi_i$ is an initial value at $t=t_i$.
The equation for $\chi$ can be expressed for $\xi=0$ 
\bea
\ddot{\chi} + 3H\dot{\chi} + (m_{\chi}^2 - m_{\phi}^2) \chi \simeq 0,
\label{vector_eom3}
\eea
where we have assumed $\frac{3}{2} m_{\chi}^2 \chi^2 
\ll \frac{1}{2}m_{\phi}^2 \phi^2$ from the slow-roll conditions ($|\dot{H}|
\ll H^2$) and  $ \chi^2 \ll m_{pl}^2$ which can be derived from
the condition $\rho_{\chi} \ll \rho_{\phi}$.
If we set $m_{eff}^2 = m_{\chi}^2 - m_{\phi}^2$, 
then  the solution of $\chi$ depends on $m_{eff}^2$.
If $\ddot{\chi}$ in (\ref{vector_eom3}) can be neglected,
 $\chi(t)$ is constant for $m_{eff}^2 = 0$ during
an inflationary period (see Fig. \ref{fig1b}). And
if $m_{eff}^2 > 0$, we can obtain
\bea
\chi(t) = \chi_i \left(\frac{\phi(t)}{\phi_i}\right)^{m_{eff}^2/m_{\phi}^2}
\approx \chi_i\left(1- \frac{m_{\phi}}{\phi_i/m_{pl}}t\right)^{m_{eff}^2/m_{\phi}^2}.
\eea
As the universe expands, $\chi(t)$ decreases.
On the contrary, if $m_{eff}^2 < 0$, then (\ref{vector_eom3}) becomes
\bea
3H\dot{\chi} - |m_{eff}^2| \chi \simeq  0,
\eea
and its solution is given by
\bea
\chi(t) = \chi_i \left(\frac{\phi_i}{\phi(t)}\right)^{-|m_{eff}^2|/m_{\phi}^2}
\approx  \chi_i\left(1- \frac{m_{\phi}}{\phi_i/m_{pl}}t\right)^{-|m_{eff}^2|/m_{\phi}^2}.
\eea
Contrary to $m_{eff}^2 > 0$ case, $\chi(t)$ increases as the universe 
expands. 

In order to check these analytic results, we compute numerically
the evolutions of $\phi$ and $\chi$. In Fig. \ref{fig1b},
we plot the evolution of $\chi$ depending on $m_{\chi}/m_{\phi}$.
We set to $m_\phi = 10^{-5} m_{pl}$, $\phi_i = 3 m_{pl}$ and $\chi_i
=10^{-3} m_{pl}$ for the computation. The evolutions of $\chi$ show 
much different behavior depending on $m_{\chi}/m_{\phi}$. The numerical
results
are consistent with the analytic results. For $m_{\chi} = m_{\phi}$,
$\chi$ shows constant behavior during an inflation period and
 after the end of inflation it begins to oscillate as a scalar field does.
If $m_{\chi} = 10^{-3} m_{\phi}~ (m_{eff}^2 < 0)$, $\chi$ increases
slowly as time goes on and starts oscillation after the end of inflation.
Finally, for $m_{\chi}  = 3 m_{\phi}~ (m_{eff}^2 > 0)$, $\chi$ 
is decreasing.

\section{Linear perturbations} \label{pert}

The linear perturbation calculations with a vector field is not
an easy task. Even if we begin with either an anisotropic spacetime
\cite{Kanno:2008gn} 
or an isotropic FRW spacetime  
using a triplet of orthogonal vectors 
\cite{ArmendarizPicon:2004pm, Golovnev:2008hv}
for a spacelike vector field, the non-vanishing background vector 
fields will make it impossible to use decomposition theorem
in the linear perturbations. This means there exist mode couplings
between scalar, vector and tensor perturbations. 
In spite of these problems,
 gravitational wave spectrum is calculated in \cite{Golovnev:2008hv}, 
in which the coupling 
terms are eliminated using the underlying symmetry
and the metric perturbations for a  spacelike vector field
are discussed in anisotropic
spacetime with the fixed-norm condition in \cite{Dulaney:2008ph}.
The power spectrum  \cite{Dimopoulos:2006ms} 
and non-Gaussianity \cite{Dimopoulos:2008yv} 
of the longitudinal and transverse component of a spacelike vector field
are calculated 
without considering the metric perturbations. 

Although  mode coupling problems  do not
arise any more
if we begin with a timelike vector field, the calculations are too messy
and complicated. 
Linear perturbations of a timelike vector field is discussed 
in \cite{Lim:2004js}. They also considered the metric perturbations
with the fixed-norm constraint.
We will discuss about the linear perturbations
without taking into account 
the metric perturbations in this section and we will try to
investigate the gravitational metric perturbations in the forthcoming
paper \cite{koh09}.

\subsection{Linear perturbations without metric perturbations}

We decompose the perturbation of the vector field into 
the scalar and vector mode perturbations
\begin{eqnarray}
\delta A_{\mu}(t,{\bf x}) =  (\delta A_0(t,{\bf x}),\delta A_i(t,{\bf x}))
\equiv  (\delta \chi, a \nabla_i \psi + a S_i),
\end{eqnarray}
where $\nabla_i S^i = 0$. Here $\delta \chi$ and $\psi$ are
scalars and $S_i$ is a vector perturbation.

\begin{figure}[t]
\includegraphics[width=0.6\textwidth]{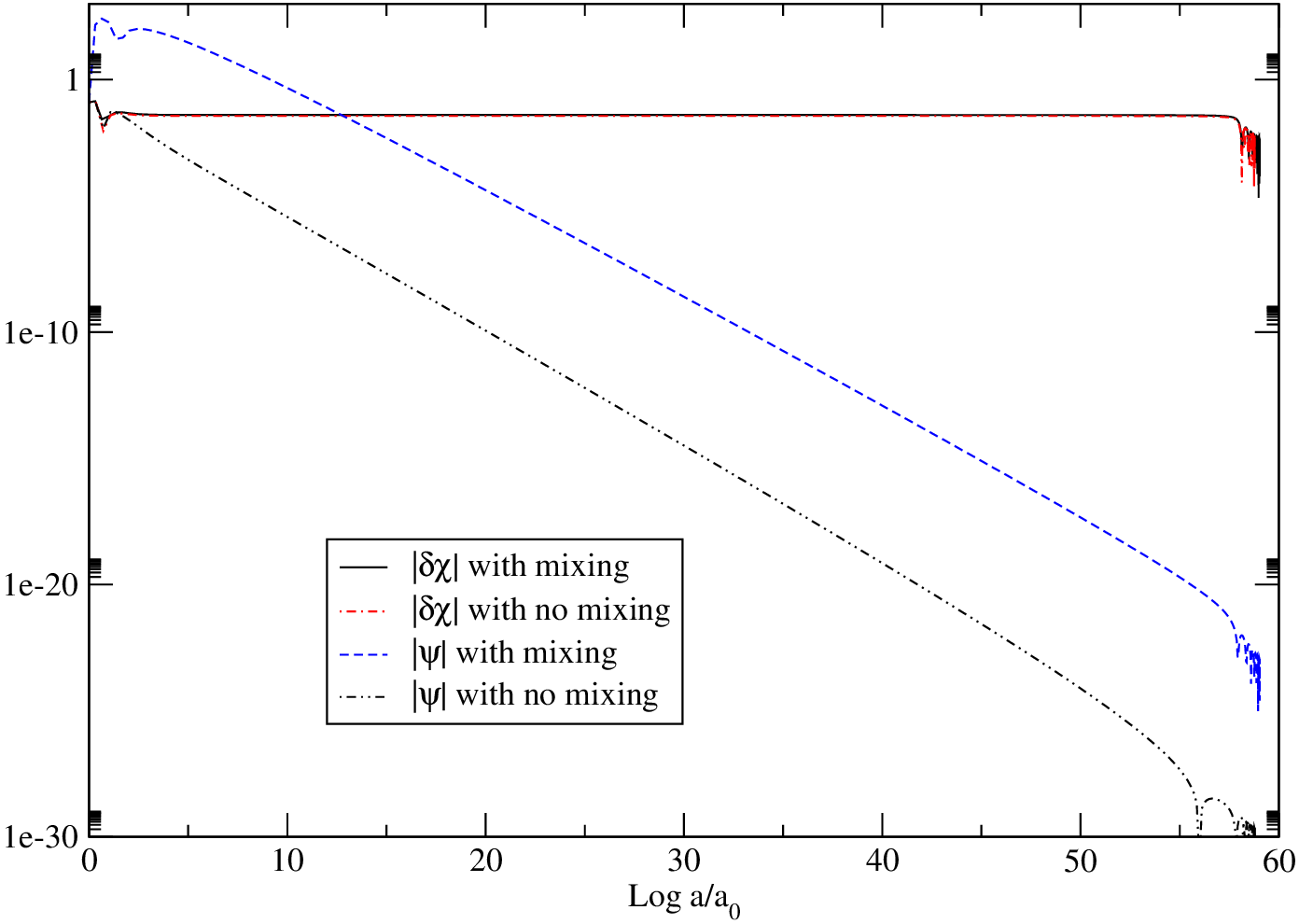}
\caption{We plot the evolutions of 
$|\delta \chi|$ and $|\psi|$ when $\xi=0$.
We have used $m_{\phi}/m_{pl} = 10^{-5}, m_{\chi}/m_{\phi}=1, \phi_i = 
3m_{pl}$ and $\chi_i = 10^{-3}m_{pl}$.}
\label{fig2a}
\end{figure}

\begin{figure}[t]
\includegraphics[width=0.6\textwidth]{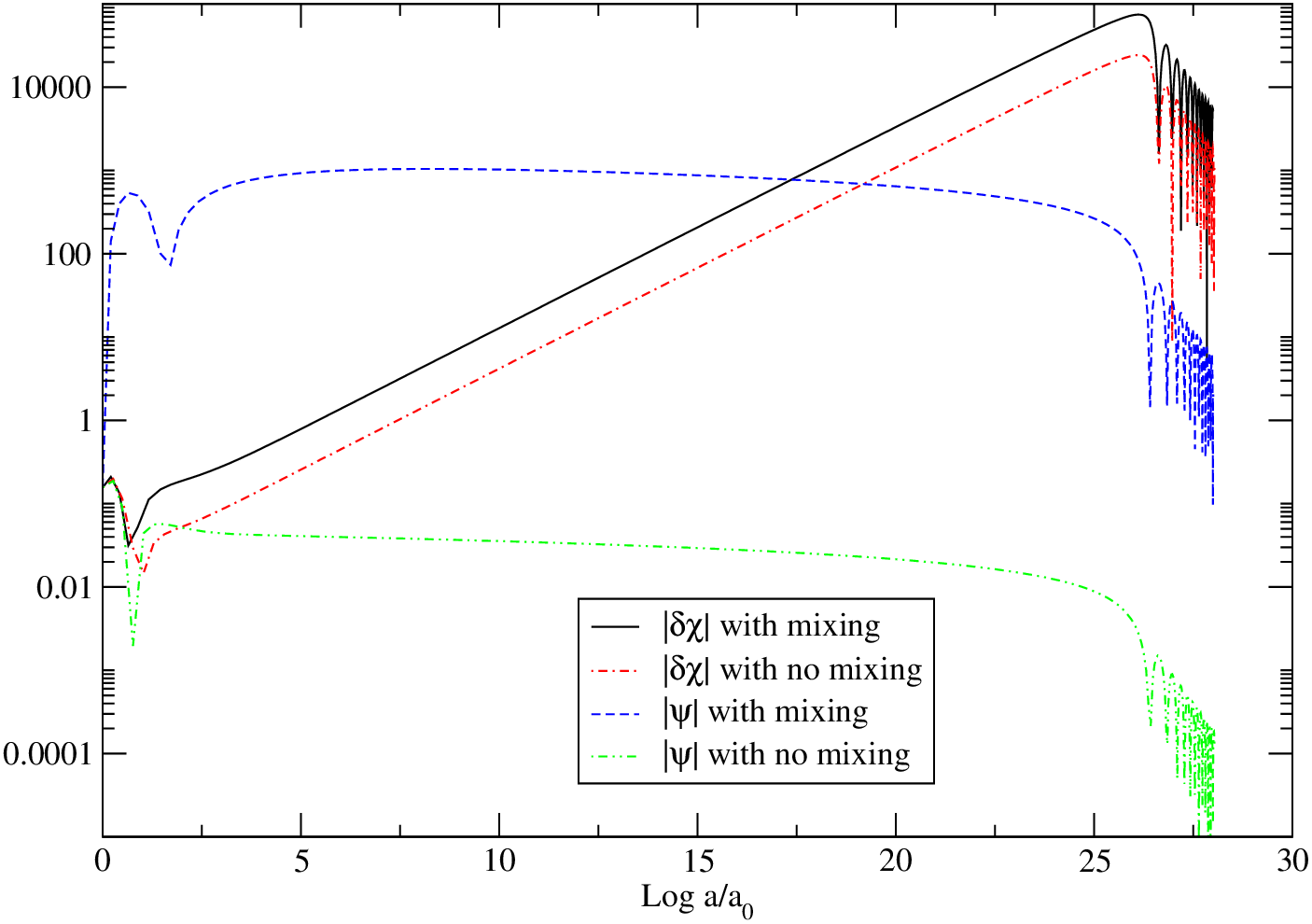}
\caption{The evolutions of 
$|\delta \chi|$ and $|\psi|$ are plotted when $\xi = 1/6$
with the same parameters as in Fig. \ref{fig2a}.}
\label{fig2b}
\end{figure}

With this decomposition of the linearized vector field,
we obtain the perturbed equations of motion in
momentum space by linearizing (\ref{eom})
\bea
& &\delta \ddot{\chi}_k +3H\delta \dot{\chi}_k 
+\left[\frac{k^2}{a^2} + 3(1-2\xi)\dot{H} -12 \xi H^2 
+m_{\chi}^2 \right] \delta \chi_k 
-2aH \frac{k^2}{a^2} \psi_k = 0,  
\label{pert_chi}  \\
& & \ddot{\psi}_k + 3H\dot{\psi}_k + 
\left[\frac{k^2}{a^2} + (1-6\xi)\dot{H} + 2(1-6\xi)H^2 
+m_{\chi}^2 \right]\psi_k
+ \frac{2H}{a}\delta \chi_k = 0, 
\label{pert_psi}\\
& & \ddot{S}_{ik} + 3H\dot{S}_{ik} + \left[\frac{k^2}{a^2} 
+ (1-6\xi)\dot{H} + 2(1-6\xi)H^2 
+m_{\chi}^2 \right] S_{ik} = 0,
\label{pert_s}
\eea
where we have used the Fourier transform
\bea
\delta \chi (t,{\bf x}) = \int\frac{d^3k}{(2\pi)^{3/2}}
\delta \chi_{k}(t) e^{i{\bf k}\cdot {\bf x}},
\eea
and similarly for $\psi$ and $S_i$.
Unlike a scalar field case,  the perturbed equations of $\delta \chi$
and $\psi$
are coupled each other.
Since it is not easy to calculate above equations analytically because of
the coupling terms,
first we calculate numerically  and the results are shown in
Figs. \ref{fig2a} and \ref{fig2b} when $k= 5 aH$. 
We also calculate the  equations for comparison
with and without the coupling terms.

We have used $m_{\phi} = 10^{-5}m_{pl},~
\phi_i = 3m_{pl},~ \chi_i = 10^{-3}m_{pl}$ and $m_{\phi} = m_{\chi}$.
We take the initial conditions for $\delta \chi_{ki}$ and $\psi_{ki}$ as
\bea
\delta \chi_{ki}, \psi_{ki} = \frac{a_i^{-1}}{\sqrt{2k}} \exp{(ik/a_i H_i)},
\eea
where the subscript $i$ denotes the initial value at $a=a_i$.
Because the equation of $S_{ik}$ has the same form with that of $\psi_{k}$ 
except the coupling term, the behavior of $S_{ik}$ is similar to 
that of $\psi_{k}$.

In Fig. \ref{fig2a}, the evolutions 
of $\delta \chi_{k}$ and $\psi_{k}$ are shown for the minimal coupling case
($\xi =0$).  In this diagram 
$\delta \chi_{k}$ shows constant behavior on super-Hubble
scales. But $\psi_{k}$ as well as $S_{ik}$ is decaying  when
the modes are larger than the Hubble horizon.  
On the contrary,
for the nonminimal coupling ($\xi = 1/6$) in Fig. \ref{fig2b}, 
the situations are
reverse. While $\psi_{k}$ and $S_{ik}$ show the constant solutions
on super-Hubble scales, $\delta \chi_{k}$ is increasing as the universe
expands.
And the amplitude of 
$|\psi|$ is enhanced due to
the effect of the mixing term in the sub-Hubble scale region.
The mixing term in $\psi$ seems to 
cause the instability in the small scale.

The coupled equations can be solved by use of the 
appropriate Green functions
\bea
\delta \chi_k(t) &=& \delta \tilde{\chi}_k(t) + \Delta \chi_k,
\label{eq:tot_chi} \\
\psi_k(t) &=& \tilde{\psi}_k(t) +\Delta \psi_k
\label{eq:tot_psi}
\eea
where $\delta \tilde{\chi}_k$ and $\tilde{\psi}_k$ are the 
solutions of Eqs. (\ref{pert_chi}) and (\ref{pert_psi}) 
in the absence of the coupling terms and
\bea
\Delta \chi_k(t)  &=& \int_{t_i}^t dt^{\prime} 
G_{\chi,k}^{R}(t,t^{\prime}) 2aH\frac{k^2}{a^2}
\tilde{\psi}_k(t^{\prime}), 
\label{eq:corr_chi} \\
\Delta \psi_k(t) &=& 
\int_{t_i}^t dt^{\prime} G_{\psi,k}^{R}(t, t^{\prime})\frac{2H}{a}
\delta \tilde{\chi}_k(t^{\prime}). 
\label{eq:corr_psi}
\eea
Here $G_{\chi,k}^{(R)}(t, t^{\prime})$ and $G_{\psi,t}^{(R)}(t,t^{\prime})$ are
the retarded Greens' functions which are defined by \cite{Birrell:1982ix}
\bea
G_{\chi,k}^{R}(t,t^{\prime}) = \theta(t-t^{\prime}) i 
(\delta \tilde{\chi}_k^{(+)}(t)\delta \tilde{\chi}_k^{(-)}(t^{\prime}) 
-\delta \tilde{\chi}_k^{(-)}(t)\delta \tilde{\chi}_k^{(+)}(t^{\prime})),
\eea
where $\theta(t-t^{\prime}) = 1$ if $t>t^{\prime}$ and $\theta(t-t^{\prime}) = 0$ 
if $t< t^{\prime}$
and similarly for $G_{\psi,k}^{R}$. Here superscripts $+$ and $-$ denote 
the positive and negative frequency mode, respectively.
These Greens' functions obey the differential equations
\bea
& &\left[\partial_t^2 + 3H\partial_t + \frac{k^2}{a^2}-12\xi H^2 
+ m_{\chi}^2\right]
G_{\chi,k}^{R} (t,t^{\prime}) = -\delta(t-t^{\prime}), \\
& &\left[\partial_t^2 + 3H\partial_t + \frac{k^2}{a^2}+
2(1-6\xi)H^2 + m_{\chi}^2\right] G_{\chi,k}^{R}(t,t^{\prime}) = -\delta (t-t^{\prime}).
\eea

First, we will try to analyze the numerical results analytically on 
super-Hubble  scales in the absence of the coupling terms.
The temporal behaviors  of the linearized vector field
perturbations even with the coupling terms can be expected to show  similar 
results on super-Hubble scales from Figs. 3 and 4.
If we change the variable $t$ into the
scale factor $a$, then the equations (\ref{pert_chi}) and
(\ref{pert_psi}) can be expressed in the large scale limit as
\bea
& &\frac{d^2 \delta \tilde{\chi}_k}{da^2} +\frac{4}{a}
\frac{d\delta \tilde{\chi}_k}{da}
- \left(12\xi - \frac{m_{\chi}^2}{H^2}\right) \frac{\delta 
\tilde{\chi}_k}{a^2} 
\simeq 0, \\
& &\frac{d^2 \tilde{\psi}_k}{da^2} +\frac{4}{a}\frac{d \tilde{\psi}_k}{da}
+\left(2(1-6\xi) + \frac{m_{\chi}^2}{H^2}\right)\frac{\tilde{\psi}_k}{a^2} 
\simeq 0,
\eea
where we have assumed $H \simeq {\rm const.}$ and $|\dot{H}|\ll H^2$.
We can obtain the following solutions
\bea
\delta \tilde{\chi}_k \sim C_k a^{p_{+}} + D_k a^{p_{-}}, \label{sol_chi} \\
\tilde{\psi}_k \sim C^{\prime}_k  a^{q_{+}} + D^{\prime}_k a^{q_{-}},
\label{sol_psi}
\eea
where
\bea
p_{\pm} = -\frac{3}{2} \pm \sqrt{\frac{9}{4}+ 12\xi - \frac{m_{\chi}^2}{H^2}},
\quad q_{\pm} =  -\frac{3}{2} \pm 
\sqrt{\frac{1}{4}+12\xi - \frac{m_{\chi}^2}{H^2}}
\eea
and $C_k,D_k, C_k^{\prime}$ and $D_k^{\prime}$ are the constant coefficients 
depending on $k$.

For $\xi =0$, if $m_{\chi} \ll H^2$, the dominant mode solution of
$\delta \tilde{\chi}_k$ is constant. But $\tilde{\psi}_k$ 
shows decaying solutions (See Fig. {\ref{fig2a}).
On the contrary,  for $\xi = 1/6$, while $\delta \tilde{\chi}_k 
\propto a^{1/2},$
the dominant mode solution of $\tilde{\psi}_k$ 
is constant (See Fig. \ref{fig2b}).

Next we calculate the power spectrum of $\delta \chi_{k}$, 
$\psi_{k}$ and
$S_{ik}$.
In order to calculate the power spectrum, we need the exact solutions
of (\ref{pert_chi}), (\ref{pert_psi}) and (\ref{pert_s}).
If we use $|\dot{H}|\ll H^2$ and $H \simeq {\rm const.}$ during
inflation, $\delta \tilde{\chi}, \tilde{\psi}$ 
and $S_i$ have the following exact form of the
solution in the absence of the coupling terms:
\bea
\delta \tilde{\chi}_k(t) &=& \left(\frac{k}{aH}\right)^{3/2}
\left[c_{1k} H^{(1)}_{\nu_{\chi}}\left(\frac{k}{aH}\right) 
+ d_{1k} H^{(2)}_{\nu_{\chi}}
\left(\frac{k}{aH}\right)\right], 
\label{esol_chi}\\
\tilde{\psi}_k(t) &=& \left(\frac{k}{aH}\right)^{3/2}
\left[c_{2k} H^{(1)}_{\nu_{\psi}}\left(\frac{k}{aH}\right) 
+ d_{2k} H^{(2)}_{\nu_{\psi}}
\left(\frac{k}{aH}\right)\right], 
\label{esol_psi}\\
S_{ik} (t) &=& \left(\frac{k}{aH}\right)^{3/2}
\left[c_{3k} H^{(1)}_{\nu_{\psi}}\left(\frac{k}{aH}\right) 
+ d_{3k} H^{(2)}_{\nu_{\psi}}
\left(\frac{k}{aH}\right)\right],
\eea
where $H^{(1)}_{\nu}(x)$ and $H^{(2)}_{\nu}(x)$ are the Hankel function
of the first and second kind, respectively, and
\bea
\nu_{\chi} = \sqrt{\frac{9}{4} +12\xi - \frac{m_{\chi}^2}{H^2}},
\quad \nu_{\psi} = \sqrt{\frac{1}{4}+ 12\xi - \frac{m_{\chi}^2}{H^2}}.
\label{nu}
\eea

In order to determine the coefficients $c_{ik}$ and $d_{ik}$,
we need initial conditions when the modes are well within the horizon,
$k/aH \rightarrow \infty$. Although quantum field theory is not
well constructed for the ghost fields, we assume in this paper 
the initial conditions for the vector field satisfy the
WKB-type solution
\bea
\lim_{k\eta \rightarrow -\infty} \delta \chi_k \approx
\frac{a^{-1}}{\sqrt{2\omega_k}} e^{-i\int^{\eta} \omega d\eta^{\prime}}
\eea
and similarly for $\psi_k$ and $S_{ik}$. 
 When 
the modes stay well inside of horizon, $\omega$ can be
approximated as $\omega \approx k$,then
\bea
\lim_{k\eta \rightarrow -\infty}\delta \chi_k \approx 
\frac{a^{-1}}{\sqrt{2k}}\exp{(-ik\eta)}.
\label{initial}
\eea

Here $\eta$ is a 
conformal time, $dt = a d\eta$, and for $H = {\rm const.}$, 
$\eta = -\frac{1}{aH}$. Using the asymptotic form of
the Hankel functions in the limit $x\gg 1$
\bea
H^{(1,2)}_{\nu}(x) \sim \sqrt{\frac{2}{\pi x}}\exp\left[\pm i (x
-\left(\nu+\frac{1}{2}\right)\frac{\pi}{2}\right],
\eea
we choose $d_{ik} = 0$ for a positive frequency mode and determine
$c_{ik}$ through the matching to the initial condition (\ref{initial}) 
\bea
c_{1k} = \sqrt{\frac{\pi}{4}}e^{i\pi(\nu_{\chi}+1/2)/2}\frac{H}{k^{3/2}}, \quad
c_{2k} = c_{3k} = \sqrt{\frac{\pi}{4}}e^{i\pi(\nu_{\psi}+1/2)/2}\frac{H}{k^{3/2}}.
\label{coeff}
\eea

With these coefficients,
 $\delta \tilde{\chi}_k$ can be expressed  in the large scale limit ($k \ll aH$)
as 
\bea
\delta \tilde{\chi}_k \simeq \sqrt{\frac{\pi}{a^3 H}}e^{i\pi/4}
\frac{e^{i\nu_{\chi} \pi/2}}{1-e^{2i\nu_{\chi}\pi}}
\left[\frac{1}{\Gamma(1+\nu_{\chi})}\left(\frac{k}{2aH}\right)^{\nu_{\chi}}
-\frac{e^{i\nu_{\chi}\pi}}{\Gamma(1-\nu_{\chi})}
\left(\frac{k}{2aH}\right)^{-\nu_{\chi}}\right]
\label{large_sol}
\eea
and $\tilde{\psi}_k$ and $S_{ik}$ are also obtained by replacing $\nu_{\chi}$
with $\nu_{\psi}$. 
Here we use the asymptotic form of $H_{\nu}^{(1)}(x)$ for $x \ll 1$
\bea
H_{\nu}^{(1)}(x) \sim \frac{2}{1-e^{2i\nu\pi}}
\left[\frac{1}{\Gamma(1+\nu)}\left(\frac{x}{2}\right)^{\nu}
-\frac{e^{i\nu\pi}}{\Gamma(1-\nu)}\left(\frac{x}{2}\right)^{-\nu}
\right].
\eea
Second term in (\ref{large_sol}) becomes a dominant mode
and first term is a subdominant mode. These solutions are
exactly consistent with the large scale solutions in (\ref{sol_chi})
and(\ref{sol_psi}).

The power spectrum for $\delta \tilde{\chi}_k, \tilde{\psi}_k$ and $S_{ik}$ are
calculated for the dominant mode 
\bea
\tilde{\calp}_{\chi}(k) &=& \frac{k^3}{2\pi^2}
|\delta \tilde{\chi}_k|^2 = \frac{4\pi \csc^2 \nu_{\chi}\pi}{\Gamma^2(1-\nu_{\chi})}
\left(\frac{H}{2\pi}\right)^2 \left(\frac{k}{2aH}\right)^{3-2\nu_{\chi}},
\label{power2_chi} \\
\tilde{\calp}_{\psi}(k) &=& \frac{4\pi \csc^2 \nu_{\psi}\pi}{\Gamma^2(1-\nu_{\psi})}
\left(\frac{H}{2\pi}\right)^2 \left(\frac{k}{2aH}\right)^{3-2\nu_{\psi}}
\label{power2_psi}
\eea
and the power spectrum of $S_{ik}$ is same as that of $\tilde{\psi}_{k}$.


The spectral indexes
for $\delta \tilde{\chi}_k$ and $\tilde{\psi}_k$ 
at late times ($k \ll aH$) are
\bea
n_{\chi} -1 &\equiv& \frac{d\ln \tilde{\calp}_{\chi}}{d\ln k}
= 3 - 2\nu_{\chi},  \\
 n_{\psi}-1 &=& 3- 2\nu_{\psi}.
\eea
From (\ref{nu}), if $m_{\chi}^2 \ll H^2$ and $\xi = 0$,
$n_{\chi} = 1$ and $n_{\psi} = 3$. But if $m_{\chi}^2 \ll H^2$ and
$\xi = 1/6$,  $n_{\chi} \simeq -0.1$ and $n_{\psi} = 1$.
In other words, for a light vector field with $m_{\chi}^2 \ll H^2$,
$\delta \tilde{\chi}_k$ gives a scale invariant spectrum only when
$\xi =0$, but both $\tilde{\psi}_k$ and $S_{ik}$ 
are scale invariant only the nonminimal
coupling case ($\xi = 1/6$).
These results are a little different from those in 
\cite{Dimopoulos:2006ms} in which 
the scale invariant spectrum for the longitudinal perturbations ($\psi_k$) 
is only possible when $\xi = 0$ 
but the transverse perturbation ($S_{ik}$)
gives a scale invariant spectrum only when $\xi = 1/6$.

Now we can calculate the power spectrum of $\delta \chi$ 
using (\ref{eq:tot_chi})
\bea
\calp_{\chi}(k) = \frac{k3}{2\pi^2}|\delta \chi_k|^2 = 
\tilde{\calp}_{\chi}(k)\left(1+
\frac{\delta \calp_{\chi}}{\tilde{\calp}_{\chi}}\right),
\label{eq:tot_pow}
\eea
and similarly for $\psi_k$ using (\ref{eq:tot_psi}). The second term
in the last expression represents the effect of the coupling term.

Although it is difficult to compute (\ref{eq:tot_pow}),
since we are interested in the amplitude of the perturbations 
well after horizon exit, 
the change in the power spectrum due to  the coupling terms
can be calculated in the limit 
$t\rightarrow \infty$  for $\xi =0$:
\bea
\frac{\delta \calp_{\chi}}{\tilde{\calp}_{\chi}}(\infty) \simeq
\frac{2\delta \tilde{\chi}_k {\rm Im}\Delta \chi_k (\infty)}
{|\delta \tilde{\chi}_k(\infty)|^2}, \quad
\frac{\delta \calp_{\psi}}{\tilde{\calp}_{\psi}}(\infty) \simeq
\frac{2\tilde{\psi}_k {\rm Re}\Delta \psi_k(\infty)}
{|\tilde{\psi}_k(\infty)|^2}.
\eea
The leading order in $\delta \tilde{\chi}$ and $\tilde{\psi}$ can be
obtained from
(\ref{esol_chi}) and (\ref{esol_psi}) in the late time limit:
\bea
\delta \tilde{\chi}_k(\infty) = i\frac{H}{\sqrt{2}k^{3/2}},
\quad
\tilde{\psi}_k(\infty) = \frac{a^{-1}}{\sqrt{2}k^{1/2}},
\eea
and $\Delta \chi_k(\infty)$ and $\Delta \psi_k(\infty)$ can be calculated  
from (\ref{eq:corr_chi}) and (\ref{eq:corr_psi})
\bea
\Delta \chi_k (\infty)&=& \frac{\pi^{3/2}2^{\nu_{\chi}-1}}{\Gamma(1-\nu_{\chi})
\sin \nu_{\chi}\pi}e^{i\pi(\nu_{\psi}+1/2)/2}\frac{H^4}{k^{7/2}}
\int_0^{x_i}dx^{\prime} x^{3/2-\nu_{\chi}}{x^{\prime}}^3
J_{\nu_{\chi}}(x^{\prime})H_{\nu_{\psi}}^{(1)}(x^{\prime}), \\
\Delta \psi_k (\infty) &=& \frac{-\pi^{3/2}2^{\nu_{\psi}-1}}{\Gamma(1-\nu_{\psi})
\sin \nu_{\psi}\pi}e^{i\pi(\nu_{\chi}+1/2)/2}\frac{H^4}{k^{11/2}}
\int_0^{x_i} dx^{\prime} x^{3/2-\nu_{\psi}} {x^{\prime}}^3
J_{\nu_{\psi}}(x^{\prime}) H_{\nu_{\chi}}^{(1)}(x^{\prime}),
\eea
where $x=\frac{k}{aH}$ and $J_{\nu}(x)$ is the Bessel function of the 
first kind.  In the limit $k \rightarrow 0$ ($x_i \rightarrow 0$), 
the leading order terms
become
\bea
{\rm Im}\Delta \chi_k(\infty) \propto \frac{k^{5/2}}{a_i^6 H^2},
\quad {\rm Re}\Delta \psi_k(\infty)  \propto \frac{k^{3/2}}{a a_i^6H^3}.
\eea
and then the leading order contributions in the change of the
power spectrum are
\bea
\frac{\delta \calp_{\chi}}{\tilde{\calp}_{\chi}}(\infty) \propto
\frac{k^4}{a_i^6 H^3}, \quad
\frac{\delta \calp_{\psi}}{\tilde{\calp}_{\psi}}(\infty) \propto
\frac{k^2}{a_i^6 H^3}.
\eea
The amplitude of the deviations
from the power spectrum of $\delta \tilde{\chi}$ 
 increases as $k^4$
for small $k$, while the deviations from the power spectrum 
of $\tilde{\psi}$ increases as $k^2$.

For $\xi =1/6$, the deviations in the power spectrum due to
the coupling terms can be calculated 
in the limit $k\rightarrow 0$,
then
\bea
& &\frac{\delta \calp_{\chi}}{\tilde{\calp}_{\chi}}(\infty) \simeq 
\frac{2 \delta \tilde{\chi}_k \Delta \chi_k(\infty)}
{|\delta \tilde{\chi}_k|^2} 
\propto \frac{k^{1/2+\nu_{\chi}}}{a_i^{\nu_{\chi}+5/2}H^{\nu_{\chi}-1/2}} , \\
& &\frac{\delta \calp_{\psi}}{\tilde{\calp}_{\psi}}(\infty) \simeq
\frac{2\tilde{\psi}_k {\rm Im}\Delta \psi_k(\infty)}{|\tilde{\psi}_k|^2}
\propto \frac{k^{3/2-\nu_{\chi}}}{a_i^{11/2-\nu_{\chi}}H^{5/2-\nu_{\chi}}},
\eea
where we have used 
\bea
\delta \tilde{\chi}_k(\infty) \propto \frac{1}{a^{3/2}H^{1/2}}
\left(\frac{k}{aH}\right)^{-\nu_{\chi}} ,\quad 
\tilde{\psi}_k(\infty) = i\frac{H}{\sqrt{2}k^{3/2}},
\eea
and the leading order terms of $\Delta\chi_k(\infty)$ and 
${\rm Im}\Delta\psi_k(\infty)$ are
\bea
\Delta \chi_k(\infty)\propto  
\frac{k^{1/2}}{a_i^{\nu_{\chi}+5/2}a^{3/2-\nu_{\chi}}}, \quad
{\rm Im}\Delta \psi_k(\infty) \propto 
\frac{k^{-\nu_{\chi}}}{a_i^{-\nu_{\chi}+11/2} H^{-\nu_{\chi}+3/2}}.
\eea
Since $\nu_{\chi} = \sqrt{17}/2 \approx 2$, 
the deviation of $\delta \tilde{\chi}$
increases as $k^{1/2+\nu_{\chi}}$ and that of $\tilde{\psi}$ decreases
as $1/k^{\nu_{\chi}-3/2}$. 

\subsection{Brief discussions about the metric perturbations}

We will briefly discuss about the linear perturbations of the
vector field including the metric perturbations in this section.
It is convenient to use a conformal time, $\eta$,  to treat
the metric perturbations, so in this section we will use the conformal time.

We consider the perturbed metric 
in scalar and vector longitudinal gauge 
\cite{Battefeld:2004cd, ArmendarizPicon:2004pm} in which
the metric takes the form 
\begin{equation}
ds^2 = a^2(\eta) \left[-(1+2\Phi)d\eta^2 -2 B_i d\eta dx^i 
+(1-2\Psi)\gamma_{ij}dx^i dx^j \right],
\end{equation} 
where $\nabla_i B^i = 0$.

With this metric, we can derive the  perturbed 
equations of motion of the time component of the vector field by
linearizing (\ref{eom})
\begin{eqnarray}
& &\delta \chi^{\prime\prime}  + 2\mathcal{H}\delta \chi^{\prime}
 - 3 \left(\mathcal{H}^2 - \mathcal{H}^{\prime}
-\frac{1}{3}(m^2-\xi R) a^2 \right)\delta \chi
-\nabla^2 \delta \chi
+2 \mathcal{H}\nabla^2 \psi \nonumber \\
& & = \chi \Phi^{\prime\prime} 
+3\chi \Psi^{\prime\prime}
+ 3\left(\chi^{\prime}-\mathcal{H}\chi\right)
(\Phi^{\prime}+\Psi^{\prime}) +8\mathcal{H}\chi \Phi^{\prime}
+2 \left(\chi^{\prime\prime} +2 \mathcal{H}\chi^{\prime} - 3\mathcal{H}^2 \chi
+3\mathcal{H}^{\prime}\chi \right)\Phi  \nonumber \\
& &~~~ +\xi a^2 \chi \delta R,
\end{eqnarray}
where $\calh = a^{\prime}/a$ and a prime denotes the derivative with respect
to the conformal time,
and  of the longitudinal (scalar) 
mode for $\nu =i$
\begin{eqnarray}
& & \psi^{\prime\prime} +2\mathcal{H}\psi^{\prime}
+\left[\mathcal{H}^{\prime}
+\mathcal{H}^2
+a^2 (m^2 -\xi R)\right] \psi 
-\nabla^2 \psi 
+ 2 \mathcal{H}\delta \chi
\nonumber \\
& & = \chi(\Phi^{\prime} +\Psi^{\prime})
+ 2\chi \Psi^{\prime}
+\left(2 \chi^{\prime} 
+ 6\mathcal{H}\chi\right) \Phi,
\end{eqnarray}
and finally of transverse (vector)
 mode for $\nu=i$
\begin{eqnarray}
 S_i^{\prime\prime} + 2\mathcal{H} S_i^{\prime} + 
\left[\mathcal{H}^{\prime}
+\mathcal{H}^2 
+a^2(m^2 -\xi R) \right] S_i - \nabla^2 S_i
 = 0.
\label{metric_eoms}
\end{eqnarray}
Since the spatial components of the background vector field vanish,
the mode couplings between different perturbation modes do not occur.
And the perturbed equation of the transverse components ($S_i$),
(\ref{metric_eoms}), is completely
decoupled from the metric perturbations. This implies that 
the power spectrum of $S_i$ at late times becomes  as in (\ref{power2_psi})
\bea
\calp_{S}(k) &=& \frac{4\pi \csc^2 \nu_{\psi}\pi}{\Gamma^2(1-\nu_{\psi})}
\left(\frac{H}{2\pi}\right)^2 \left(\frac{k}{2aH}\right)^{3-2\nu_{\psi}},
\eea
 where $\nu_{\psi}$ is given in (\ref{nu}),
and hence $S_i$ is scale invariant when the light vector field is
nonminimally coupled to gravity ($\xi =1/6$)
even the metric perturbations are taken into account.

\section{Conclusions} \label{conclusion}

We have investigated the dynamics of a vector field, which violates 
Lorentz invariance, during an accelerating phase in the early universe.
In order to avoid a spatial anisotropic background, 
we employ the timelike vector 
field. We have shown that the timelike vector field 
is difficult to realize successful inflation since
the effective mass of the  vector field is order of the Hubble scale
and the slow-roll conditions could not be fulfilled.
Contrary to the spacelike vector field inflation model \cite{Golovnev:2008cf}
in which slow-roll phase can be realized by
introducing a nonminimal coupling, it turns out that 
the nonminimal coupling
does not help to generate an accelerating phase.

Although the timelike vector field could not generate inflation,
we would expect it to play a role as a curvaton after the end of inflation 
which is driven by a scalar field. So we calculated the evolution of
the vector field during a scalar field driven inflation period. The vector
field can roll down slowly enough if the mass of the vector field 
is similar to that of a scalar field. 
We need to calculate the linear perturbations
in order to check if the vector field can generate 
a scale invariant power spectrum.

Although we only consider the non-vanishing 
time component of the vector field on the
spatially isotropic background spacetime 
to avoid anisotropy, the spatial component
perturbation  as well as the 
time component perturbations of the vector field
should be taken into account. The spatial component perturbations can be
decomposed into the longitudinal and transverse part. 
The time component perturbation and the longitudinal
mode of the spatial component perturbation are coupled each other and  
these coupling terms cause the instability in sub-Hubble scale.
It would be necessary to investigate the instability due to
the coupling terms in small scale \cite{koh09}.

If the transverse
perturbations can give a scale invariant spectrum, it may be responsible
for some large scale anomaly \cite{Eriksen:2003db} in CMB power spectrum. 
We have calculated
the power spectrum and spectral indexes 
of the time and spatial component perturbations of
the vector field without 
considering the gravitational metric perturbations. 
In the absence of the coupling terms,
the time component perturbation of the vector field gives
a scale invariant spectrum when $\xi = 0$, but the longitudinal and
transverse perturbation of the vector field provide a scale invariant
spectrum when the vector field is coupled nonminimally to  gravity.
The fact that
both longitudinal and transverse perturbations have a scale
invariant spectrum only when $\xi =1/6$ is different from 
the results of \cite{Dimopoulos:2006ms} in which 
the  longitudinal perturbation has a  scale invariant spectrum when $\xi = 0$
but the transverse perturbation does when $\xi = 1/6$. 
 Further, we have calculated 
the amplitude of the 
deviations from the power spectrum which are calculated in the
absence of the coupling terms.
The deviations of $\delta \chi$ rise as $k^{4}$ when $\xi=0$ and
$k^{5/2}$ when $\xi=1/6$ for small $k$, while the deviations of
$\psi$ rise as $k^2$ when $\xi=0$ and fall like $1/k^{1/2}$ when $\xi=1/6$.

Note that even if we consider the gravitational metric perturbations,
the transverse perturbation of the vector field is not affected by the
metric perturbations as shown at the end of Section \ref{pert}.
So the transverse perturbation of the vector field still gives a
scale invariant spectrum only when $\xi = 1/6$. 
But it would be necessary to calculate the power spectrum and spectral indexes
of the time component and longitudinal perturbation of
the vector field with including the metric perturbations.

\acknowledgments
We would like to thank R. Cai for useful discussions. HB would like to
thank Xian Gao for useful discussions. SK would like to
appreciate the hospitality of the Center for Quantum
Space Time (CQUeST) at Sogang University. 
BH was supported in part by a grant from the Chinese Academy of Sciences with
No. KJCX3-SYW-N2, grants from NSFC with No. 10821504 and No. 10525060.

\end{document}